\documentclass{aastex}

\slugcomment{Submitted to the Astronomical Journal}

\shorttitle{Quintessence \& Radio Galaxies}
\shortauthors{Daly \& Guerra}

\begin{document}


\title{Quintessence, Cosmology, and FRIIb Radio Galaxies}


\author{Ruth A. Daly}
\affil{Department of Physics, Penn State University, P. O. Box 7009, Reading, 
    PA, 19610-6009}
\email{rdaly@psu.edu}

\and

\author{Erick J. Guerra}
\affil{Department of Chemistry and Physics, Rowan University, 
201 Mullica Hill Road, Glassboro, NJ 08028-1701}
\email{guerra@scherzo.rowan.edu}



\begin{abstract}  
FRIIb radio galaxies provide a modified standard yardstick 
that allows constraints to be placed on 
global cosmological parameters. This modified
standard yardstick is analogous to the 
modified standard candle provided by 
type Ia supernovae.  
The radio galaxy and supernova 
methods provide a measure
of the coordinate distance to high-redshift sources, and the 
coordinate distance is a function of global cosmological parameters. 

A sample of 20 FRIIb radio galaxies with 
redshifts between zero and two are compared with the parent population
of 70 radio galaxies to determine the coordinate distance to 
each source.  The coordinate distance determinations
are used to 
constrain the current mean mass-energy density of quintessence 
$\Omega_Q$, the equation of state
of the quintessence $w$, and the current mean mass-energy 
density of non-relativistic matter $\Omega_m$; zero  
space curvature is assumed.  Radio galaxies alone indicate that the 
the universe is currently accelerating in its expansion (with 84\% 
confidence); most of the allowed 
parameter space falls within 
the accelerating universe 
region on the $\Omega_m - w$ plane.  This provides  
verification of the acceleration of the universe 
indicated by high-redshift supernovae, and suggests
that neither method is plagued by systematic errors.
It is 
found that $\Omega_m$ must be less than about 0.5 
and the equation
of state $w$ of the quintessence must lie between 
-0.25 and -2.5 at about 90\% confidence.

Fits of the radio galaxy data constrain the model
parameter $\beta$, which describes a relation between the beam
power of the AGN and the total energy 
expelled through large-scale jets.  It is shown that 
the empirically determined model parameter 
is consistent with
models in which the outflow results from
the electromagnetic extraction of rotational energy
from the central compact object. A specific relation
between the strength of the magnetic field near the AGN, 
and the spin angular momentum
per unit mass of the central compact object is predicted.   

\end{abstract}


\keywords{cosmology: cosmological parameters --- cosmology: observations --- 
cosmology: theory --- cosmology: dark matter --- galaxies: active}


\section{Introduction}

There are several independent 
ways to determine the global cosmological
parameters that describe the current state of the universe.
It is important to have 
several complementary and independent methods that yield 
consistent results since any given method could be plagued
by unknown systematic errors.  A 
particularly useful way to determine 
global cosmological parameters is through measurements 
of the coordinate distance to high-redshift sources.  This 
method is particularly useful because the coordinate
distance depends only upon global cosmological parameters,
and is independent of 
the clustering properties of the mass-energy
components that control the
expansion rate of the universe, as long as each component
is homogeneous and isotropic on very large scales.  
The coordinate distance is also independent of 
whether different components cluster differently, 
known as biasing, and on whether the dark matter
is cold, warm, or hot (though it does depend on
the equation of state of each component).  

Two cosmological tools that are 
particularly sensitive to the coordinate distance as a function of
redshift are FRIIb radio galaxies, which provide a modified standard
yardstick (Daly 1994, 2002a; 
Guerra 1997; Guerra \& Daly 1998 [GD98]; Guerra, Daly, 
\& Wan 2000 [GDW00]), and type Ia supernovae, which provide a modified standard
candle (e.g. Riess et al. 1998, 
Perlmutter et al. 1999).  Some aspects of the methods
are compared in \S \ref{compare}.  

The coordinate distance to a source at a given redshift depends on 
the present value of the mean mass-energy density of each 
component, and on the redshift evolution of the mass-energy
density of each
component.  Non-relativistic 
matter, including baryonic matter and clustered dark matter, have 
a mean mass-energy density that evolves as $(1+z)^3$.  
There is a contribution
from radiation and neutrinos left over from the big bang,
currently negligible, which 
has an energy-density that 
evolves as $(1+z)^4$.  There may
also be a cosmological constant, which constant energy
density, and so evolves as $(1+z)^0$.  

The redshift evolution of the mean mass-energy density of a  
component depends on the equation of state $w = P/\rho$ of the 
component. There could exist 
a dynamical vaccum energy density called Quintessence 
(Caldwell, Dave, \& Steinhardt 1998).
Quintessence would have an equation of state $w$ such that the density
evolves as $(1+z)^n$, where $n=3(w+1)$ (see Turner \& White 1997; 
Bludman \& Roos 2001, Wang et al. 2000; or the 
Appendix of this paper).    
For example, pressure-less dust (i.e., non-relativistic matter) 
is described by $P=w=0$ and $n=3$; a relativistic
fluid is described by $w = 1/3$ and $n=4$, and 
a cosmological constant is described by 
$w= -1$ and $n=0$.  

Two components of the universe are known to exist with 
certainty: there is a non-relativistic component (including
baryons and clustered dark matter),
with zero-redshift, normalized mean mass-energy density 
$\Omega_m$, and a 
primordial relativistic component that is negligible for
the epochs of interest here (including microwave
background photons and 
neutrinos).  

There must be at least one more component, which 
remains to be determined.  This 
third component could be a cosmological constant, 
quintessence, space
curvature, or something else.  In the simplest case, there is only one
additional component that is significant and has a measurable
impact over the redshift interval where the
coordinate distance is used to constrain cosmological parameters
and the properties of the unknown component.  

Here, the determination of the coordinate distances
to 20 radio galaxies with redshifts between zero and two are used
to constrain the equation of state and current mean mass-energy
density of quintessence assuming a spatially flat 
universe as indicated by measurements of
the microwave background radiation 
(de Bernardis et al. 2000, Balbi et al. 2000).

The radio galaxy method is briefly reviewed in \S \ref{radgals}. 
Constraints on cosmological parameters and the equation
of state of quintessence 
obtained using FRIIb radio 
galaxies as a modified standard yardstick
are presented in \S \ref{quint}.    
The 
supernovae and radio galaxy methods and results
are compared in detail
in \S \ref{compare}.  The relation between beam 
power and total energy for FRIIb radio jets is discussed in \S
\ref{beta}. Here it is shown that the underlying
hypotheses of the radio galaxy method is consistent with 
current models of large-scale jet production in AGN.  
The results are summarized in 
\S \ref{summary}.

\section{Cosmology FRIIb Radio Galaxies} \label{radgals}

The use of FRIIb radio galaxies as a modified 
standard yardstic to determine global cosmological parameters
has been presented in detail elsewhere; see Daly (1994, 2002a), GD98, and 
GDW00.  Radio images of the 20 sources used here can be found
in Leahy, Muxlow, \& Stephens (1989), Liu, Pooley, \&
Riley (1992), and GDW00.
The radio galaxy method is 
based on the following premises.  First, it is assumed that the
forward region of the radio bridge of an FRIIb source
represents a strong shock
front, and the equations of strong shock physics are applied at
this boundary.  Second, it is assumed that  
there is a power-law relation between the average beam power
of the source $L_j$ and the total lifetime of the outflow
$t_*$: $t_* \propto L_j^{-\beta/3}$ 
(see \S \ref{beta}).
Third, it is assumed that, at 
a given redshift, the sources studied will have an average source
size similar to that of the parent population of FRIIb sources
at the same redshift, where the source size is defined as the separation
between the radio hot spots.  All of these points are discussed
in detail in previous work (e.g. Daly 2002a,b; GDW00).

In addition, in its simplest interpretation, the model 
works in a straight-forward way if the radio power
of a given FRIIb radio galaxy does not 
decrease monotonically and measurably with time.  
There is substantial evidence that this is the case for radio galaxies
that fall into the FRIIb category, as do all of the radio galaxies
considered here (Daly 2002b).  For example, Neeser et al. (1995) find
that there is no correlation between radio size and radio power.  
The low frequency radio surface brightness of these sources can
be explained using adiabatic expansion alone if the radio powers
of the sources are roughly time-independent (Wellman, Daly, \& 
Wan 1997).  Gopal-Krishna, Kulkarni, \& Wiita (1996) find that
a roughly constant radio power followed by an exponential drop 
in radio power would explain the relative sizes of radio galaxies
and radio loud quasars in the context of the orientation unified
model.  And, it is shown in \S \ref{beta} that the beam power is
independent of core-hot spot separation, which would not
be expected if the radio power of a given source decreases
with time.  Thus, there is no empirical evidence that the
radio power of a given FRIIb radio galaxy decreases with time,
and substantial evidence that it is roughly constant over the
lifetime of a given source, followed by a precipitous drop.
The Mach numbers of FRIIb radio galaxies are about 2.5 to 10
(Wellman, Daly, \& Wan 1997), implying
overpressures of 10 to 200, so the lobe will rapidly expand
and drop in radio power when the beam power shuts off, 
as suggested by Neeser et al. (1995).   
Some authors (e.g., Blundell, Rawlings, \& Willcot 1999) 
have studied large heterogeneous samples that 
include sources with all
types of radio structure and power and both
radio loud quasars and radio galaxies.  
To explain
the properties of this heterogenous sample, a
model is proposed that allows for the decrease in the 
radio power of a given source as a function of time.  There is,
however, no evidence that this applies to the select subset
of sources (the most powerful FRII radio galaxies)
studied here, and substantial evidence that
the radio powers of FRIIb radio galaxies do not monotonically
decline over the lifetime of a given source.  

Since FRIIb radio galaxies form a remarkably homogenous population,
and all the sources at a given redshift are quite similar,
the average size of a given source 
$D_*$ will equal the average size
of the full population $<D>$ of FRIIb radio galaxies
with similar redshifts. The dependence of
$D_*$ and $<D>$ on the coordinate distance, and hence
on cosmological parameters, is quite different, and
\begin{equation}
<D>/D_* \propto (a_or)^{2\beta/3 + 3/7}
\end{equation} 
since $<D> \propto (a_or)$ 
(see GD98 eq. 6).  By construction, the ratio
$<D>/D_*$ should be a constant, independent of redshift.
Because the model is based upon a comparison of individual 
source characteristics with the characteristics of the parent
population at the same redshift, selection effects, such as
radio power selection effects, are unlikely to affect the
constraints placed on cosmological parameters.  All of
the sources for which values of $D_*$ are determined
belong to the parent population.  The sources for which
$D_*$ is determined are subject to selection effects
that are identical to those of the parent population,
used to determine $<D>$.  Thus, this method is one 
of the few methods that is unlikely to be affected 
by observational selection effects.  Equation (1)
allows constraints to be placed on 
the model parameter $\beta$, and the cosmological parameters
that enter into the determination of the coordinate
distance.

\section{Constraints on Quintessence From Radio Galaxies} \label{quint}

Recent measurements of the cosmic microwave background anisotropy suggest
that the universe has zero space curvature 
(de Bernardis et al. 2000, Balbi et al. 2000).  Here,
a universe with zero space curvature ($k=0$) and quintessence is
considered.  
The coordinate distance to a source at redshift $z$ 
in a spatially flat universe is 
\begin{equation}
(a_or) =  H_o^{-1} \int_{0}^{z} dz/E(z)~.
\label{eqcoor1}
\end{equation}
where $E(z) = \sqrt{\Omega_m (1+z)^3
+ (1-\Omega_m)(1+z)^n}$.  The coordinate distance is
obtained by substituting this into eq. \ref{eqcoor1} and
solving for $(a_or)$ (see, for example, 
Turner \& White 1997; Bludman \& Roos 2001;
the Appendix of this paper).  

The equation of state $w$ of quintessence is assumed to be 
time-independent over the redshift
interval from zero to two (Wang et al. 2000).  
Thus, the mean mass-energy density 
of quintessence will evolve as $(1+z)^n$, where 
$w=n/3-1$.  For a present
value of the normalized mass-energy density of quintessence $\Omega_Q$,
and that of non-relativistic matter $\Omega_m$, we have 
$\Omega_Q + \Omega_m=1$ at $z=0$, since 
space curvature is taken to be zero. This implicitly 
assumes that there is only one
unidentified component, $\Omega_Q$, that contributes significantly 
over the redshift interval where the coordinate distance will be determined.  

The minimization of the chi-squared of $<D>/D_*=$ constant leads
to constraints on $\Omega_m$, $w$,and the model parameter $\beta$.
The one and two-dimensional confidence intervals 
have been determined in the usual way
(see, for example, Press, Teukolsky, \& Flannery 1992).

\notetoeditor{Figure 1 should appear here} 
Figure 1 shows the 68\%, 90\%, and 95\% one-dimensional confidence
contours in the $\Omega_m - w$ plane.  At 90\% confidence, $\Omega_m$ 
is less than about 0.5 for any value of $w$.  A cosmological constant,
represented by $w= -1$ is clearly consistent with the radio data.
This Figure differs slightly from the two-dimensional
confidence intervals presented in Figure 2 of Daly \& Guerra 2001.
At 68 \% confidence, $\Omega_m < 0.25$, and
$-1.3 < w < -0.45$.  At 90 \% confidence,
$\Omega_m < 0.5$ and $-2.6 < w < -0.25$.
Note, some models for the dark energy invoke 
mechanisms that behave like quintessence with w less
than -1 (e.g.Frampton 2002).  

The fifty-four supernovae identified by Perlmutter et al. 
(1999) and used to obtain 
their ``primary fit'' C have been similarly fitted, and 
imply strong contraints on $\Omega_m$, though the supernovae
do not place a lower bound on w, as shown in Figure 2.

\notetoeditor{Figure 2 should appear here}

As described by Turner \& White (1997) 
and presented in some detail 
in the Appendix, the expansion rate 
of the universe is accelerating at the present epoch if 
\begin{equation} 
1+3w(1-\Omega_m) <0~.
\end{equation}  This line is drawn on the 
$\Omega_m - w$ plane (see Figures 1 and 2); points below 
the line represent solutions for which the
universe is currently accelerating, 
while those above the line represent solutions
for which the universe is currently decelerating.
Radio galaxies alone (Fig 1) indicate that the expansion of the universe is
accelerating ($q_o<0$); this is an 84\% confidence result.
 
\notetoeditor{Figure 3 should appear here} 
It is very important to consider 
whether significant covariance exists between the different 
parameters determined by the fit.   
Figure 3 shows the 68\%, 95\%, and 99\%
one-dimensional confidence contours in the $w - \beta$ plane.  Clearly, there
is no covariance between the equation of state $w$ of quintessence and
the model parameter $\beta$: $w$ and $\beta$
are independent.

\notetoeditor{Figure 4 should appear here} 
Figure 4 shows the  68\%, 95\%, and 99\% 
one-dimensional confidence contours in the $\Omega_m - \beta$ plane.  
Clearly the two parameters
are independent.  The data suggest that 
$\beta \simeq 1.75 \pm 0.25$ independent of the
cosmological parameters $\Omega_m$, $\Omega_Q$, 
or the equation of state $w$ of
quintessence.  

\section{Comparison of Supernova and Radio Galaxy Methods} \label{compare}

The observation of the acceleration of the universe through
measurements of the coordinate distance to high redshift sources
is so important that it needs to be independently verified 
using different methods.  In \S 4, it was shown that the 
radio galaxy method alone indicates at about 84\% confidence  
that the universe is accelerating 
in its expansion at the present epoch if the universe has zero
space curvature.  The supernovae teams report similar results at
stronger confidence levels (Riess et al. 1998, Perlmutter et al. 1999),
as illustrated in Figure 2.    
It is important to consider the
similarities and differences between the methods to insure that the
results of each are independent, and to compare the quantities
that are actually determined in each method.  
The two methods are compared in 
Table 1.  
 
\notetoeditor{Table 1 should appear here}

The two methods are similar is their dependence on 
the coordinate distance, and in the sense that both
provide a modified standard rather than an absolute standard.
They are somewhat similar in their redshift interval of
coverage, and in the number of sources studied.  
The radio galaxy and supernova methods 
differ in the way the ``modification'' relation is derived; one
is empirical and one is based on physical arguments.  They
also differ in that the radio galaxy method is not normalized
at zero redshift, and is independent of the properties of
local sources and of the local distance scale; 
in fact, nearly identical results obtain when
the low redshift bin is excluded from the analysis (GDW00).
The supernova relation is derived at low
redshift.  

Thus, the radio galaxy method provides independent
verification of the supernova results.  
Current constraints on cosmological parameters 
determined using the radio galaxy method 
are consistent with those obtained using the
supernova method.
Clearly, any selection effects, or unknown systematic errors, must
be completely different for these two methods.  The supernova
method involves
short-lived optical events, and the radio galaxy method involves
much more long-lived radio AGN events. The methods are derived in 
vastly different ways, and are normalized differently.  
However, similar constraints obtain, suggesting 
each method is reliable.

\subsection{Comparison Allowing for Space Curvature}

It is interesting to compare the constraints placed on cosmological
parameters assuming that three main terms control the expansion
of the universe at the present epoch: non-relativistic matter with
contribution $\Omega_m$, a cosmological constant with 
contribution $\Omega_{\Lambda}$, and space curvature. 
\notetoeditor{Figures 5 and 6 should appear here} 
Figures 5 and 6 show constraints placed by the
microwave background radiation (Bond et al. 2000), and
the high-redshift
supernovae project (Riess et al. 1998); the results obtained
by the supernovae cosmology project
(Perlmutter et al. 1999) are similar and are presented
by Daly \& Guerra (2001).  
Each
method carves out a different part of parameter space because
each method covers a different redshift range with the supernovae
being at the lowest redshift, the radio galaxies at higher
redshift, and the microwave background radiation at still
higher redshift.  The three methods are thus complementary in
their redshift coverage.  

The radio galaxy and supernovae methods are independent of the 
initial power spectrum of density fluctuations, and hence 
are independent of the index of the primordial fluctuation
spectrum.  These methods are also independent of the Hubble
constant, and of the baryon fraction.  The radio galaxy and
supernovae methods depend only on the global cosmological
parameters $\Omega_m$ and $\Omega_{\Lambda}$ allowing for 
space curvature, or $\Omega_m$ and $w$ allowing for quintessence,
where $\Omega_Q = 1-\Omega_m$, and assuming zero space curvature.
They are therefore particularly important to study since most
other methods depend on several other parameters such as the 
initial spectrum of fluctuations, the Hubble constant, and/or
the baryon fraction.  

\section{Lifetime, Beam Power, and Energy in FRIIb Radio Galaxies}\label{beta} 

The application of the radio galaxy cosmology method
indicates that $\beta \simeq 1.75 \pm 0.25$ irrespective
of cosmological model, as discussed here for
quintessence, by GDW00 in cosmologies that include
space curvature, and summarized by Daly (2002a).    
The total energy released
over the lifetime of the FRIIb in the form of a
highly collimated outflow is $E_* = L_j~t_*$ assuming
that $L_j$ is roughly constant over the lifetime
of a given source, which is supported by lack of correlation between
$L_j$ and the hotspot to core distances in these sources
(see Wan, Daly, \& Guerra 2000; Guerra \& Daly 2001).
Values of $L_j$ and $t_*$ are obtained independently,
using different aspects of the radio data 
for a given source.  Two values are determined for
each source, one on each side of each source.  
An indicator of possible problems with the model assumptions
or selection biases is whether the 
beam powers, total source lifetimes, or
total energies depend strongly on the distance
of the radio hot spot from the location of the
AGN.  Each of these quantities is plotted as a function
of hot spot distance from the AGN in 
Figures 7, 8, and 9.  There are 40 data points on
each figure, two for each of the 20 radio galaxies
studied here. 

\notetoeditor{Figures 7, 8, and 9 should appear here} 
It is interesting to note that the total energy processed
through large-scale jets  has a rather small range, 
from $\sim few \times 10^5~M_{\odot}$ to 
$\sim few \times 10^6~M_{\odot}$, though it does increase systematically with 
redshift, as shown in Figure 10.

\notetoeditor{Figure 10 should appear here}

Because $t_* \propto L_j^{-\beta/3}$ and 
$E_* = L_j~t_*$, the relation between beam power and energy released by 
an FRIIb is
$L_j \propto E_*^{3/(3-\beta)} \propto E_*^q$, where
$q = 2$ to 3 for $\beta = 1.5$ to 2.0.  In 
terms of the total lifetime of the
source, $t_* \propto L_j^{(-1/2,~-7/12,~-2/3)}$ for 
values of $\beta$ of 1.5, 1.75, and 2.0 respectively.
Clearly, $\beta$ remains far from the limiting cases
of $\beta$ = 0 or 3.  

\subsection{Electromagnetic Energy Extraction from Rotating Holes}

Blandford (1990) gives a summary of the energy and beam
power of outflows associated with the electromagnetic extraction of the
spin energy of a rotating black hole.  The beam power
and total energy available
are
\begin{equation}
L_j = L_{EM} \sim 10^{45} (a/m)^2 ~B_4^2 ~M_8^2 \hbox{ erg s}^{-1}
~\propto (a/m)^2 ~B^2~ M^2
\label{ljeq}
\end{equation}
and 
\begin{equation}
E_* = E \sim 5 \times 10^{61} (a/m)^2~M_8 \hbox{ erg}
~\propto (a/m)^2~M~,
\label{estareq}
\end{equation}
for $(a/m) <<1$, 
where $M$ is the mass of the black hole, 
$M_8$ is the mass in units of $10^8 M_{\odot}$, 
$a$ is the spin angular momentum $S$ per unit
mass $M$: $a=S/(Mc)$, c is the speed of light, 
$m$ is the gravitational radius $m=GM/c^2$, 
$B$ is the magnetic field strength, and
$B_4$ is the magnetic field strength in 
units of $10^4$ G (Blandford 1990).  

Empirically, $E_*  \propto L_j^{1-\beta/3}$
with $\beta = 1.75 \pm 0.25$.  This is consistent with
equations (4) and (5) when the magnetic field strength
satisfies
\begin{equation}
B \propto M^{(2\beta-3)/2(3-\beta)}~(a/m)^{\beta/(3-\beta)}~.
\end{equation}
This is particularly simple when $\beta = 1.5$; in 
this case $B \propto (a/m)$.  For the cases
$\beta = 1.75$ and 2, the empirical constraint is consistent
with equations (4) and (5) when
$B \propto (a/m)^{1.4}~M^{0.2}$ and 
$B \propto (a/m)^2~M^{1/2}$ respectively.  

Typical values for the beam power, total energy processed by a jet, 
and lifetime are
$L_j \sim 10^{45} \hbox{ erg s}^{-1}$, $E_* \sim 5 \times
10^5 M_{\odot} c^2$, and $t_* \sim 10^7$ yr (
see Figures 7, 8, and 9).   
Equations \ref{ljeq} and \ref{estareq} imply
a total source lifetime of  
$t_* \sim 10^9/(M_8B_4^2)$, indicating a lifetime
of about $10^7$ yr for $M_8B_4^2 \sim 10^2$.  This 
is satisfied for $M_8 \sim 10$, and $B_4 \sim 3$.  
In this case, the beam power is $\sim 10^{45} \hbox{ erg s}^{-1}$
for $(a/m) \sim (1/30)$, and the total energy is 
$\sim 5 \times 10^5 M_{\odot}$.  These values for
$M_8$, $B_4$, and $(a/m)$ seem quite reasonable.
The scaling between variables required by the empirically
determine relation between total energy and beam power
are discussed above.

\section{Conclusion} \label{summary}

Type IIb radio galaxies and type Ia supernovae are particularly
important methods to develop to constrain the global cosmological
parameters $\Omega_m$, $\Omega_{\Lambda}$, and space curvature,
or $\Omega_m$, $\Omega_Q$, and the 
equation of state of quintessence $w$.  These methods 
depend only upon the properties of global cosmological 
parameters, and 
are independent of other factors such
as the index of the primordial power spectrum, the Hubble
constant, the baryon fraction, the properties of the dark
matter that clusters around galaxies and clusters of galaxies,
any biasing of dark relative to luminous matter,  
etc.  The radio galaxy and  supernova 
methods are completely independent and have
completely different potential systematic errors, as discussed
in \S \ref{compare}.  

Radio galaxies may be used to constrain the mass-energy density and
equation of state of quintessence.  These results are presented here
assuming a spatially flat universe, which is supported by
recent measurements of fluctuations of the cosmic microwave
background.  Radio galaxies alone
suggest that the universe is accelerating in its expansion at present
(see S \ref{quint}), consistent with results obtained by the supernovae
teams, indicating that both methods are working 
well and probably are not plagued by unknown systematic
errors.  

The implications for models of energy extraction for the cases
of an outflow related to accretion and the Eddington luminosity,
and electromagnetic energy extraction of rotational energy are
considered.  If the outflow is produced by the electromagnetic
extraction of energy from a rotating black hole, then the magnetic
field strength must be related to the spin angular momentum of the
rotating black hole $S$, the mass of the black hole $M$, and the
gravitational radius of the black hole $m$, as described in
\S 5.  The relation is particularly simple if $\beta=1.5$, and
implies that the magnetic field strength satisfies $B \propto (a/m)$,
where $a =S/(Mc)$.  

\acknowledgments

We are grateful to Lin Wan and Greg Wellman who
have contributed significantly our studies of 
radio galaxies.   
We also thank Megan Donahue, 
Paddy Leahy, Chris O'Dea, Adam Riess, Max Tegmark,
and an anonymous referee  
for helpful comments.
This work was supported in part by a National Young Investigator Award 
AST-0096077 from the National
Science Foundation, and by Penn State University.
Research at Rowan University was supported in part by 
the College of Liberal Arts and Sciences and National Science Foundation
grant AST-9905652.

\clearpage

\begin{deluxetable}{crrrrrrrrrrr}
\footnotesize
\tablecaption{Comparison Between Supernova and Radio Galaxy Methods. \label{tbl-1}}
\tablewidth{0pt}
\tablehead{
\colhead{Supernovae} & \colhead{Radio Galaxies}  
}
\startdata
Type SNIa & Type FRIIb \\
$\propto (a_or)^{2.0}$ & $\propto (a_or)^{1.6}$\\
$0<z <1$ & $0<z<2$ \\
$\sim 100$ sources&20 sources (70 in parent pop.)\\
modified standard candle & modified standard yardstick\\
light curve $\Longrightarrow$ peak luminosity&radio bridge $\Longrightarrow$
average length\\
empirical relation  & 
physical relation \\
written in terms of observables&written in terms of physical variables\\
model parameter determined at $z=0$ & model parameter not determined at $z=0$\\
may depend on local distance scale& independent of local distance scale\\
universe is accelerating& $\Omega_m$ is low \\
&universe is acclelerating if k=0\\
some theoretical understanding  & 
good theoretical understanding \\
well tested empirically&needs more empirical testing\\

 \enddata


\end{deluxetable}


\clearpage



\figcaption[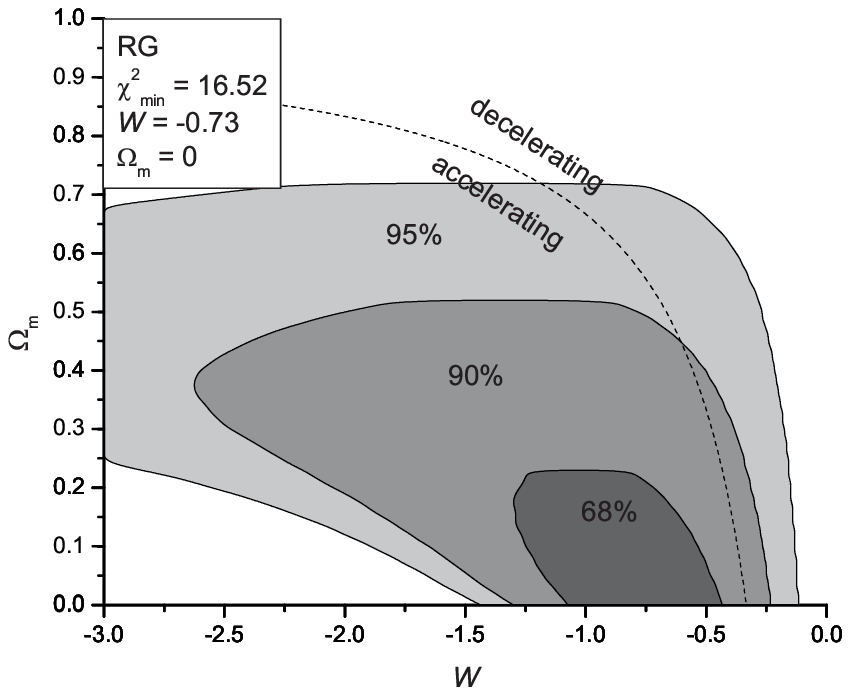]{One-dimensional confidence contours
(68\%, 90\%, and 95\%)
on the $\Omega_m - w$ plane obtained for a spatially flat
universe.  The dark dotted curve separates 
the plane into an ``accelerating'' and a ``decelerating'' 
region. The best fitting values of parameters, as well as the
chi-squared of the fit (obtained for 20 radio galaxies and
16 degrees of freedom) are indicated on the figure.
  \label{fig1}}

\figcaption[Daly.fig2.eps]{One-dimensional confidence contours
(68\%, 90\%, and 95\%)
on the $\Omega_m - w$ plane obtained for 
the 54 supernovae in the ``primary fit'' C of
Perlmutter et al. (1999) for 
a spatially flat
universe.  The dark dotted curve separates 
the plane into an ``accelerating'' and a ``decelerating'' 
region. The best fitting values of parameters, as well as the
chi-squared of the fit (obtained for 54 supernovae and
50 degrees of freedom) are indicated on the figure.
  \label{fig2}}

\figcaption[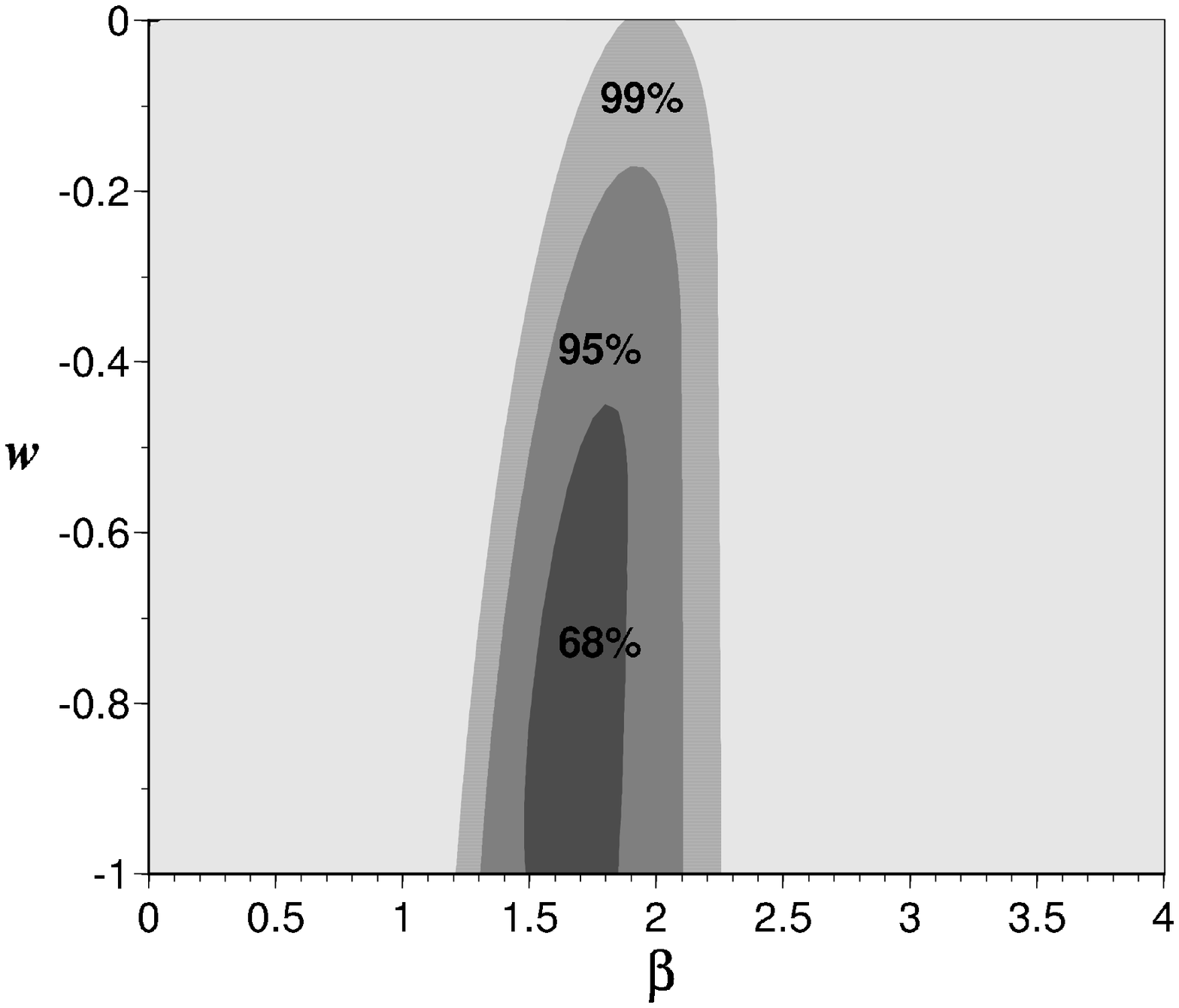]{One-dimensional confidence contours 
(68\%, 90\%, and 95\%) on the 
$w-\beta$ plane obtained allowing for 
quintessence.  
Figure shows that $\beta$ and $w$ are independent.
\label{fig3}}

\figcaption[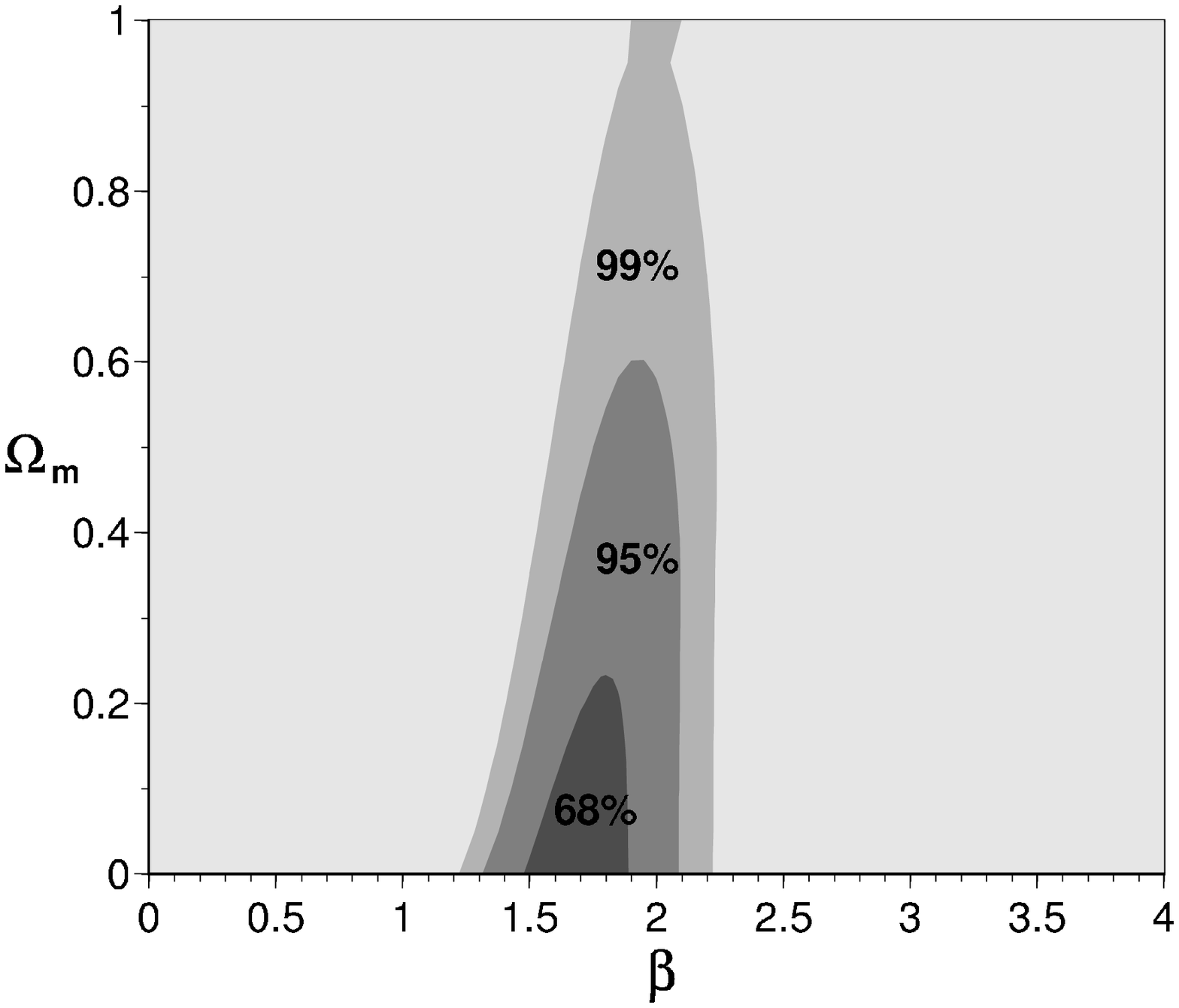]
{One-dimensional confidence contours (68\%, 90\%, and 95\%)
on the $\Omega_m - \beta$ plane, 
obtained allowing for quintessence.  
\label{fig4}}

\figcaption[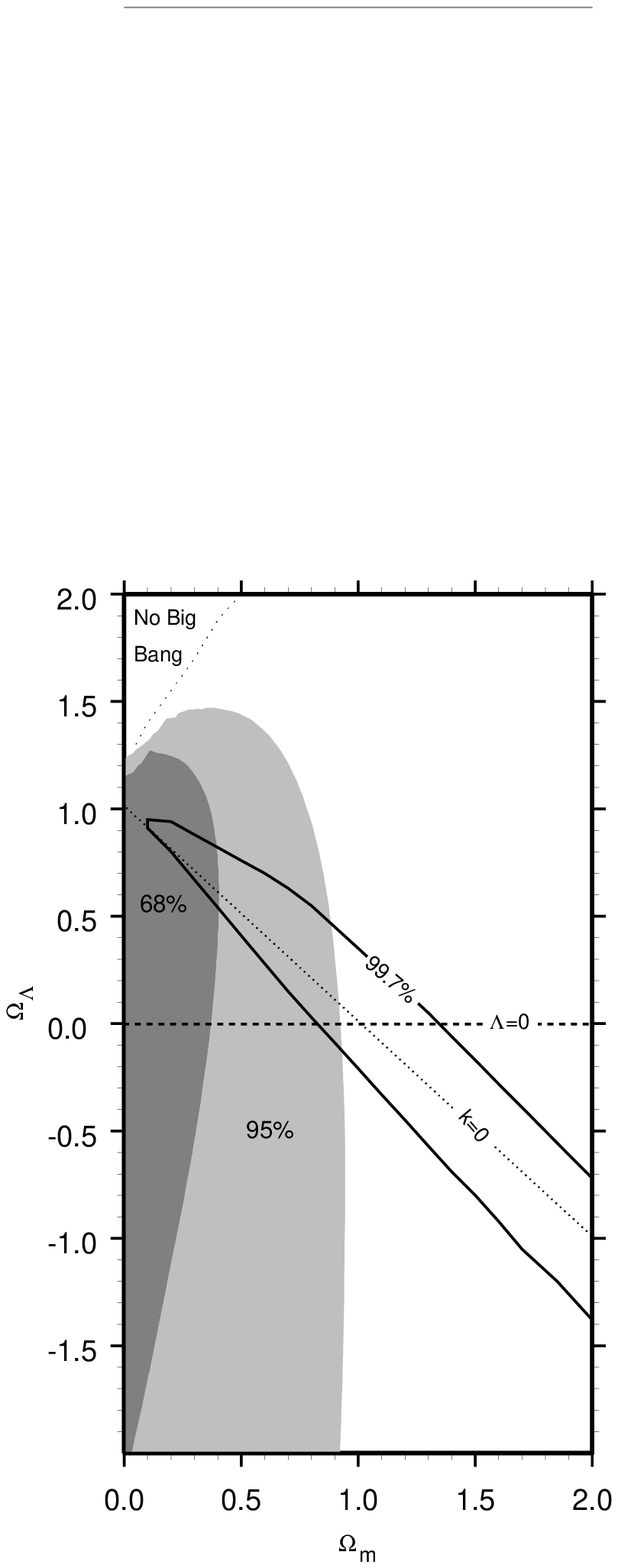]
{The two-dimensional confidence 
intervals in the $\Omega_m$-$\Omega_{\Lambda}$ plane from the cosmic
microwave background and FRIIb radio galaxies.  
Solid lines 
indicate the constraints from the cosmic microwave 
background (Bond et al. 2000)
and shaded regions indicate the constraints from FRIIb radio galaxies
(GDW00).
The combined constraints
indicate
that there must be a significant contribution from a component
other than non-relativistic matter at the present epoch.  
\label{fig5}}

\figcaption[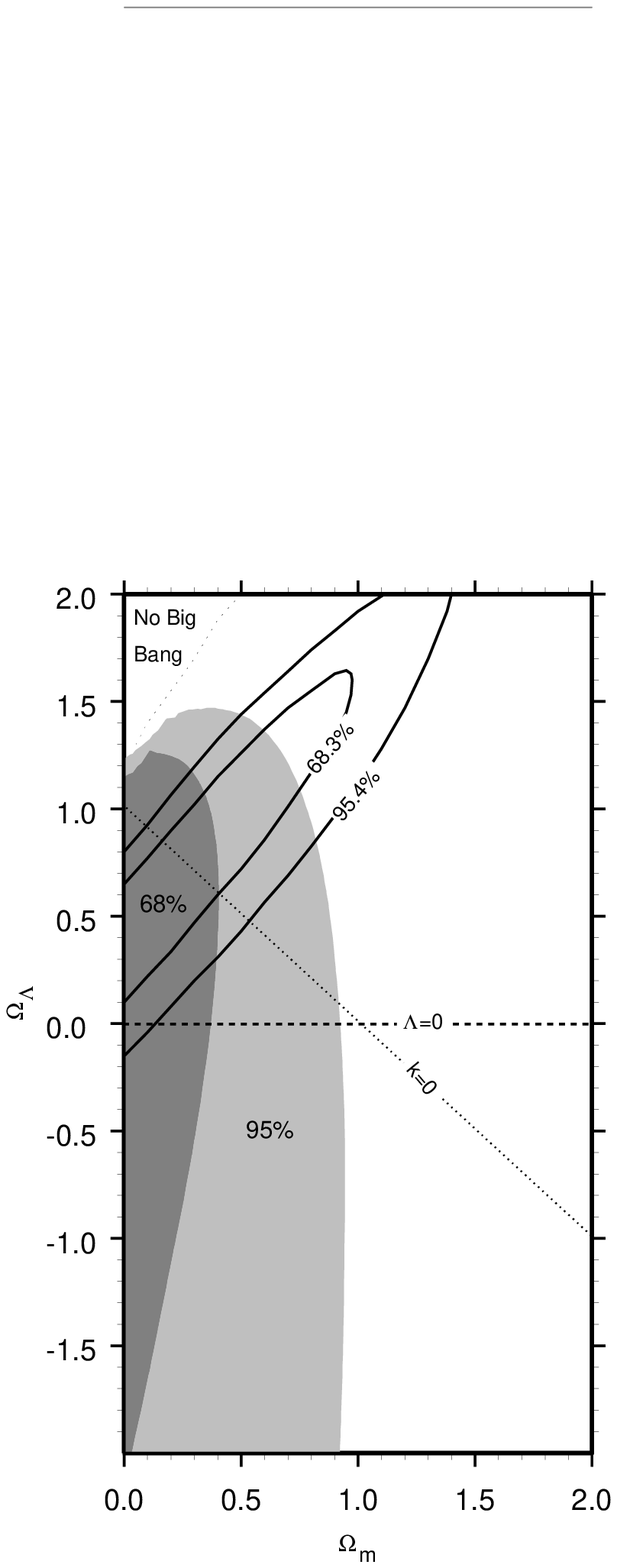]
{The two-dimensional confidence 
intervals in the $\Omega_m$-$\Omega_{\Lambda}$ plane from supernova and
FRIIb radio galaxies. 
Solid lines 
indicate the constraints from the high-redshift 
supernovae team (Riess et al. 1998)
and shaded regions indicate the constraints from FRIIb radio galaxies
(GDW00).
The 1 $\sigma$ concordance region is intersected
by the k=0 line, suggesting that the universe
is spatially flat with a low value for $\Omega_m$.  
The results obtained by the supernovae cosmology team
(Perlmutter et al. 1999) are very similar to the supernovae
results shown here.  
\label{fig6}}

\figcaption[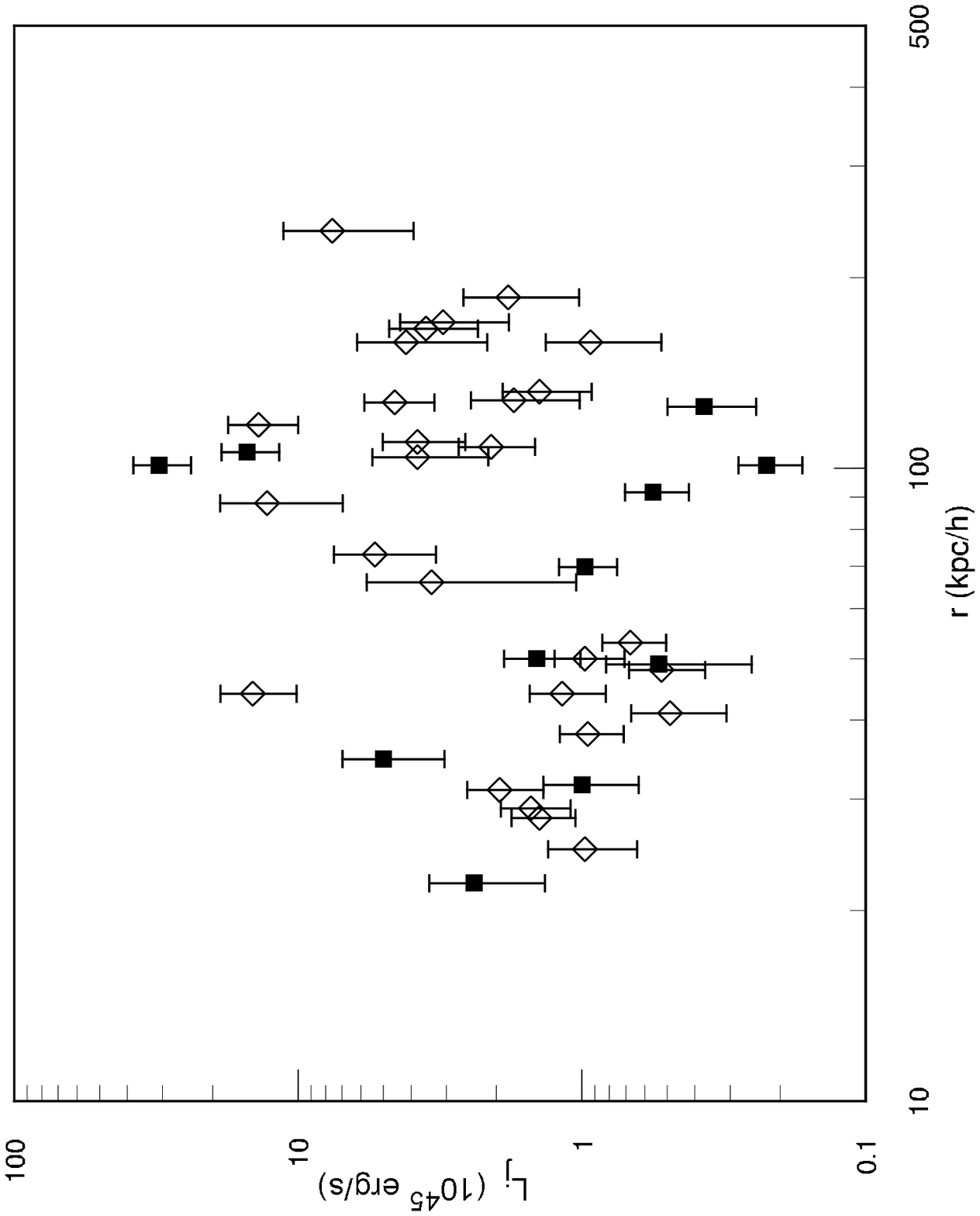]{The beam power, obtained by applying strong
shock conditions to the forward region of the radio source,
is plotted as a function of hot spot distance from the 
origin of the source.  Two values
are obtained for most sources, one for each side of each source.  
The normalization is described in WDG00 and GDW00. 
Filled symbols are the 
six new radio galaxies obtained from the VLA archive and
described in GDW00; open symbols are the 14 original radio galaxies
described in WDG00 and elsewhere. \label{fig7}}

\figcaption[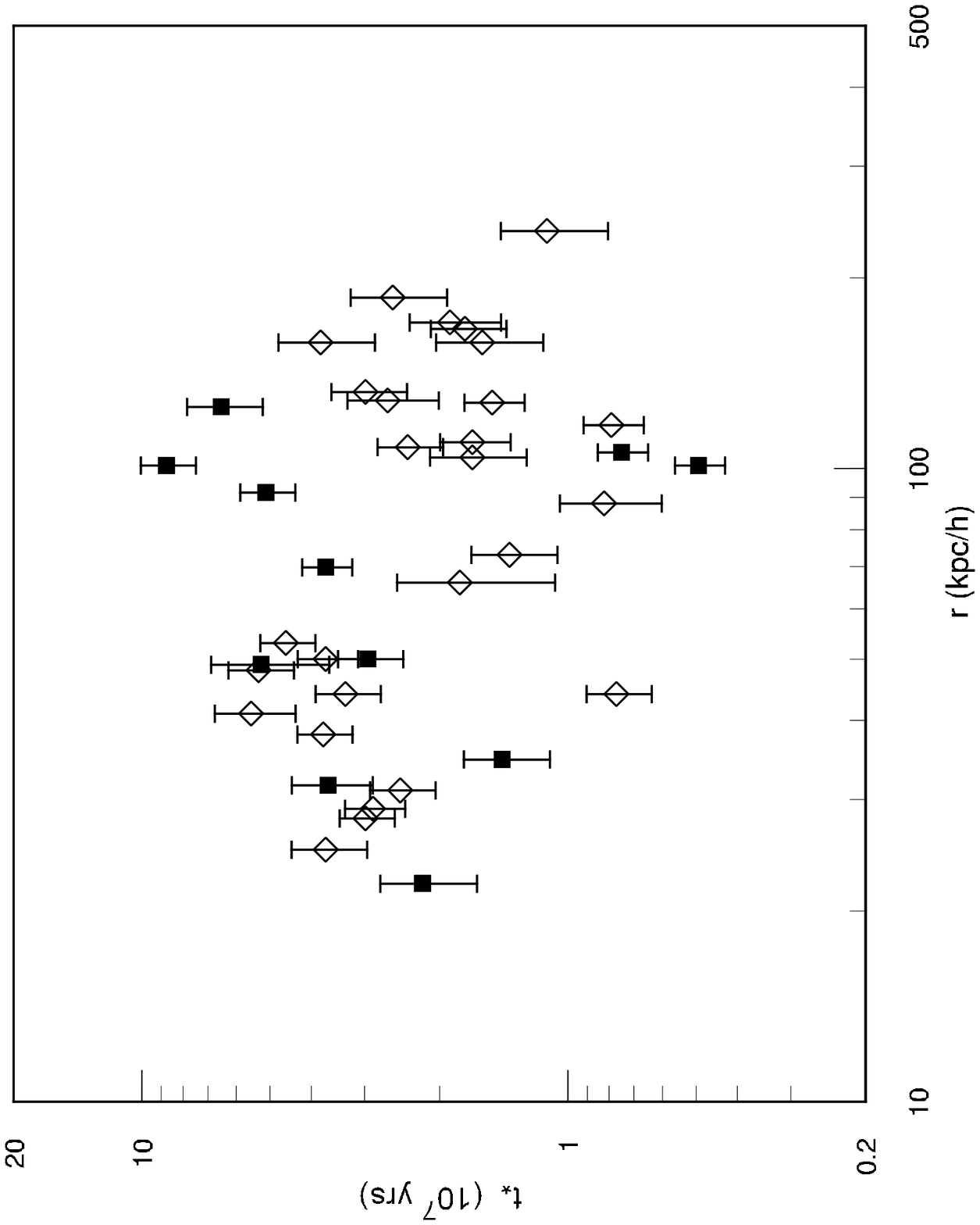]{The total time a particular AGN will produce
each of its two large-scale jets is plotted as a function of distance from 
the origin of the source.    
The filled and open symbols are the same as in figure 7.  
\label{fig8}}

\figcaption[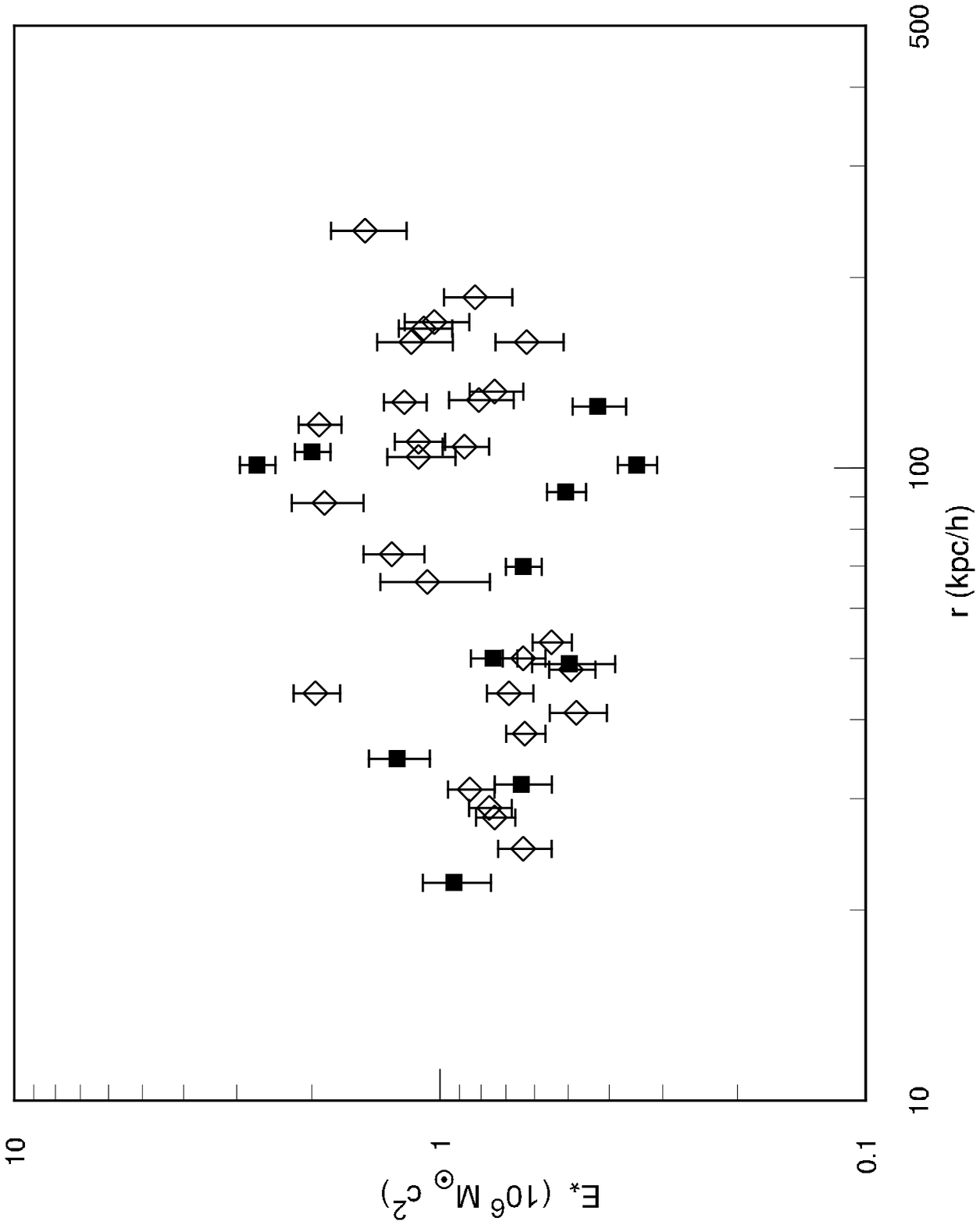]{The total energy a given AGN will process
through each of its two 
large-scale jets is plotted as a function of distance from 
the origin of the source.    
The filled and open symbols are the same as in figure 7. 
\label{fig9}}

\figcaption[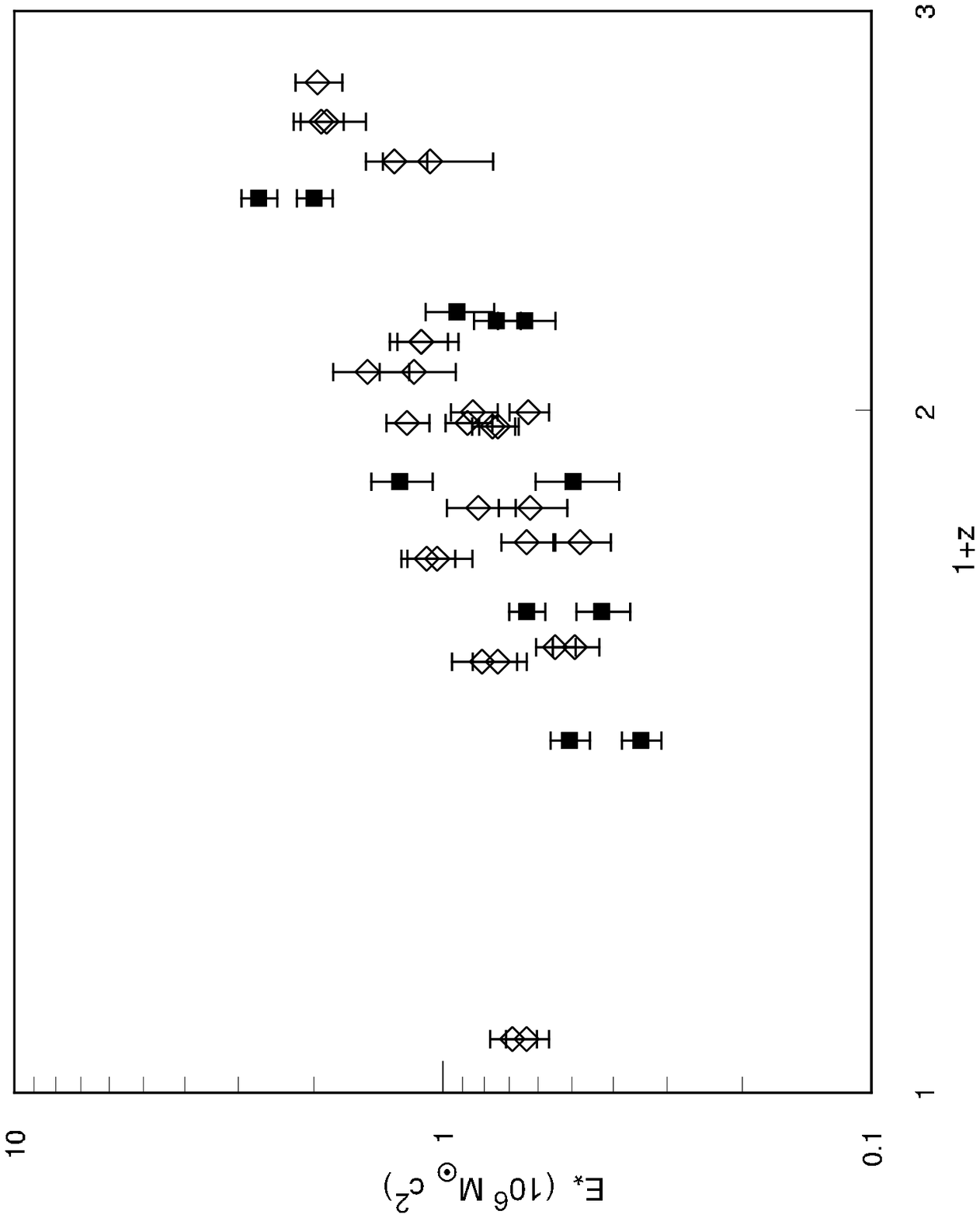]{The total energy a given AGN will process 
through each of its two large-scale jets is plotted as a function of 1+z.  
The filled and open symbols are the same as in figure 7.
\label{fig10}}

\section{Appendix} \label{Appendix}

The deceleration parameter is defined to be
$q_o=- (\ddot{a} a)/\dot{a}^2$ evaluated at
z=0, where $a$ is the cosmic
scale factor.  
This may also be written $q_o = -\ddot{a}_o/(a_o H_o^2)$, where
quantities evaluated at z=0 have a subscript 'o', and
$H_o = (\dot{a}_o / a_o)$. It is straight-forward
to show that (e.g. Peebles [1993] eq. 5.15)
\begin{equation}
(\ddot{a}/a) = -{4 \pi G \over 3}~ \sum (\rho_i + 3 p_i)
= -{4 \pi G \over 3}~ \sum \rho_i(1 + 3 w_i)
\label{acceq}
\end{equation} where $p_i$ is the pressure, $\rho_i$ is the 
mean mass-energy density, $w_i$ is the equation
of state of $i\mbox{th}$ component, $w_i = p_i /\rho_i$, and the
quintessence or cosmological constant term is included in the summation.
 
For each component, mass-energy conservation implies
\begin{equation}
\dot{\rho_i} = -3(\rho_i + p_i)(\dot{a}/a)
\label{conseq}
\end{equation}
(e.g. Peebles [1993] eqs. 5.16).
By definition, the equation of state is
$w_i=p_i/\rho_i$, so the eq. \ref{conseq} implies
$(\dot{\rho_i}/\rho_i)=-3(1+w_i)(\dot{a}/a)~.$
When the equation of state $w_i$ is time independent,
the solution to this equation is 
$\rho_i = \rho_{i,o}(1+z)^{3(1+w_i)}$, where $(1+z)=a_o/a$.  
Thus, a component with equation of state $w_i$ and
present mean mass-energy density $\rho_o$ will have 
a mean mass-energy density at redshift $z$ of
$\rho=\rho_o(1+z)^{n_i}$, where $n_i=3(1+w_i)$.   

When $k=0$, 
\begin{equation}
(\dot{a}/a)^2 = {8 \pi G \over 3}~ \sum \rho_i~,
\label{eqadot}
\end{equation}
where $\rho_i = \rho_{i,o}(1+z)^n$.  
It follows that $H_o^2 = (\dot{a}_o/a_o)^2 
= ({8 \pi G \over 3}) \sum \rho_{i,o}$.
For $k=0$, $\sum \rho_{i,o}=\rho_{c,o}$ where
$\rho_{c,o}$ is the critical density at redshift zero. Thus,
 \begin{equation}
H_o^2 = {8 \pi G \over 3}~ \rho_{c,o}
\label{Hnot}
\end{equation}
for $k=0$.
The deceleration parameter $q_o =
-\ddot{a}_o/(a_o H_o^2)$
then becomes
\begin{equation}
q_o = (1/2) \sum \Omega_i(1+3w_i)~,
\end{equation}
where $\Omega_i = \rho_{i,o}/\rho_{c,o}$ and
equations \ref{acceq} and \ref{Hnot} have
been used.  
 
The universe is accelerating rather than
decelerating when $q_o <0$.  This can 
only occur if $(1+3w_i)<0$, or 
$w_i <-1/3$, which is a necessary
but not sufficient condition to have an
accelerating universe at the present
epoch.  If there are only two significant
types of mass-energy controlling the expansion
rate of the universe at the present epoch,
quintessence and non-relativistic matter, then
the deceleration parameter is 
$q_o = \Omega_m/2 + \Omega_Q(1+3w)/2$.  
The universe will be 
accelerating in its expansion when  
\begin{equation}
1+3w(1-\Omega_m) <0~,
\end{equation}
which follows since $\Omega_Q=1-\Omega_m$.  
Thus, a curve can be drawn on the 
$\Omega_m - w$ plane; points below 
the curve represent solutions for which the
universe is currently accelerating, 
while those above the curve represent solutions
for which the universe is currently decelerating
(e.g. see Figure 1).  For example, this curve is
shown on Figure 1, which illustrates the constraints obtained using radio 
galaxies alone, and is discussed in more detail below.  Note
that radio galaxies alone place 
interesting constraints on $\Omega_m$ and
$w$, and these results are consistent with those 
obtained using other methods (e.g Wang
et al. 2000).  

The coordinate distance to a source at redshift
$z$ follows from the equation
\begin{equation}
\int dr/\sqrt{1-kr^2} = \int dt/a(t) =(1/a_o) \int (\dot{a}/a)^{-1} dz~
\label{eqcoor}
\end{equation}
(Weinberg 1972).  
For a spatially flat universe, the left hand
side of the equation reduces to the coordinate distance
$r$.  Equations
\ref{eqadot} and \ref{Hnot} imply that
$(\dot{a}/a)^2 = (H_o^2/\rho_{c,o}) \sum \rho_{i,o}(1+z)^{n_i}$
or
\begin{equation}
(\dot{a}/a)^2 = H_o^2 \sum \Omega_i (1+z)^{n_i}~.
\end{equation}
Following the notation 
of Peebles (1993), this reads $(\dot{a}/a)=H_o E(z)$, 
where 
\begin{equation}
E(z) = \sqrt{\sum \Omega_i (1+z)^{n_i}}~.
\label{eqeofz}
\end{equation}  
Thus, 
the coordinate distance to a source at redshift $z$ 
in a spatially flat universe is (see equation \ref{eqcoor}) 
\begin{equation}
(a_or) =  H_o^{-1} \int_{0}^{z} dz/E(z)~.
\label{eqcoor1}
\end{equation}

A spatially flat universe has $\sum \Omega_i = 1$.
The components that must be included in equation
\ref{eqeofz} are those that contribute from redshift 
zero out to the redshift
that the coordinate distance is being determined.  
Two components are considered here: non-relativistic matter
with normalized mean mass-energy density at $z=0$ of $\Omega_m$,
and quintessence with normalized mean mass-energy density
at $z=0$ of $\Omega_Q$.  Thus, $\Omega_m + \Omega_Q = 1$, and
equation \ref{eqeofz} becomes $E(z) = \sqrt{\Omega_m (1+z)^3
+ (1-\Omega_m)(1+z)^n}$.  The coordinate distance is
obtained by substituting this into eq. \ref{eqcoor1} and
solving for $(a_or)$.

\end{document}